\documentclass[journal=jctcce,manuscript=article]{achemso}

\usepackage{chemformula} 
\usepackage[T1]{fontenc} 
\usepackage{arydshln}
\usepackage{pgfplots}

\author{Ahmadreza Ghanbarpour}
\affiliation{Department of Medicinal Chemistry and Molecular Pharmacology, College of Pharmacy, Purdue University, 575 Stadium Mall Drive, West Lafayette, Indiana 47906, United States}
\author{Amr H. Mahmoud}
\affiliation{Department of Medicinal Chemistry and Molecular Pharmacology, College of Pharmacy, Purdue University, 575 Stadium Mall Drive, West Lafayette, Indiana 47906, United States}
\author{Markus A. Lill}
\affiliation{Department of Pharmaceutical Sciences, University of Basel, Klingelbergstrasse 50, 4056 Basel, Switzerland}
\email{markus.lill@unibas.ch}
\phone{++41 61 2076135}
\alsoaffiliation{Department of Medicinal Chemistry and Molecular Pharmacology, College of Pharmacy, Purdue University, 575 Stadium Mall Drive, West Lafayette, Indiana 47906, United States}

\title[watsiteprediction]
{On-the-fly Prediction of Protein Hydration Densities and Free Energies using Deep Learning}
 
\abbreviations{}


\begin{document}


\begin{abstract}
The calculation of thermodynamic properties of biochemical systems typically requires the use of resource-intensive molecular simulation methods. 
One example thereof is the thermodynamic profiling of hydration sites, i.e. high-probability locations for water molecules on the protein surface, which play an essential role in protein-ligand associations and must therefore be incorporated in the prediction of binding poses and affinities.
To replace time-consuming simulations in hydration site predictions, we developed two different types of deep neural-network models aiming to predict hydration site data. 
In the first approach, meshed 3D images are generated representing the interactions between certain molecular probes placed on regular 3D grids, encompassing the binding pocket, with the static protein. These molecular interaction fields are mapped to the corresponding 3D image of hydration occupancy using a neural network based on an U-Net architecture. 
In a second approach, hydration occupancy and thermodynamics were predicted point-wise using a neural network based on fully-connected layers. In addition to direct protein interaction fields, the environment of each grid point was represented using moments of a spherical harmonics expansion of the interaction properties of nearby grid points.
Application to structure-activity relationship analysis and protein-ligand pose scoring demonstrates the utility of the predicted hydration information.

\end{abstract}

\section{Introduction}
Deep learning methods have been recently used for drug discovery applications such as the prediction of binding affinity \citep{kdeep} or identifying native poses in protein-ligand docking. Ragoza et al. \citep{gnina}, for example, used 3D CNNs based on protein and ligand atomic densities to distinguish native from incorrect poses. Deep learning has also been shown to reproduce results generated by computationally demanding methods, such as prediction of molecular properties using quantum-mechanical calculations \citep{quantumml}.
Recently, machine learning methods have been deployed to reproduce simulation data or to assist MD simulations to reduce expensive computations and to improve performance in computational biology and chemistry applications \citep{mlinfrared,mlmdnature,No2019,Lubbers2018,Smith2017,vonLilienfeld2018}.

One recent application of MD simulations in drug discovery is the prediction of the positions and thermodynamic profiles of water molecules in binding sites using methods such as WaterMap \cite{watermap1,watermap2} or WATsite \cite{watsite,Yang_2017,Masters2018}. This hydration information can be used to estimate desolvation free energy contributions to a ligand's binding affinity or the potential for water-mediated interactions. To obtain this hydration information resource-intensive MD simulations are required. 

In addition of water replacement and reorganization, ligand binding is also typically associated with conformational changes of the protein \cite{Lill2011}. Recently, we demonstrated the influence of conformational protein changes on hydration site positions and thermodynamics \cite{Yang2014,Yang2016JCTC}. These studies concluded that hydration site prediction on flexible proteins needs to be performed on alternative protein states further increasing the required computational resources. Furthermore, we recently demonstrated the general importance of water networks around the bound ligand for forming enthalpically favorable complexes \cite{Lill2019}. Thus, it would be desirable to re-calculate hydration site information in an efficient manner for each bound ligand or even binding pose during docking. As this information is impossible to achieve in an efficient-enough manner with MD-simulation based techniques, modern machine learning approaches may pose an efficient alternative for predicting WATsite data on the fly, therefore eliminating the need for long, resource-exhaustive simulations.

Here, we present the first attempts for generating hydration site data on-the-fly using deep learning. Based on the MD simulations on thousands of protein structures, WATsite analysis is performed to predict hydration density and thermodynamic profiles on a 3D grid. This data is used as input to the training process utilizing two different deep learning architectures. Based on convolutional neutral networks (CNN), the first approach aims to predict hydration site information of all grid points in the binding site in a single calculation. In contrast, the second model predicts hydration information for each grid point separately using spherical-harmonics local descriptors that encode water-protein interactions at this position and the local environment of the water molecule.

\section{Material and Methods}
\subsection{Water prediction on proteins}
Several methods \cite{Nittinger2018} have been devised to identify water molecules adjacent to proteins' surfaces including knowledge-based methods such as WaterScore \citep{waterscore} or AcquaAlta \citep{aquaalta}, statistical and molecular mechanics approaches such as 3D-RISM \citep{3drism} or SZMAP \citep{szmap}, Monte-Carlo methods such as grand-canonical Monte Carlo (GCMC) simulations \citep{gcmc}, and molecular-dynamics (MD) methods such as WaterMap \citep{watermap1,watermap2} or WATsite \citep{watsite} (Figure 1). GCMC- and MD-based hydration-site prediction is accurate and widely accepted as gold-standard for computing the likely water-positions in the binding sites of proteins, its enthalpy and entropy contribution to desolvation. A recent analysis on the structure-activity relationships for different target systems demonstrated the superiority of simulation-based water prediction compared to other commercial methods such as SZMAP, WaterFLAP and 3D-RISM \cite{Bucher2018}.

\begin{figure*}
  \centering
  \includegraphics[width=16.5cm]{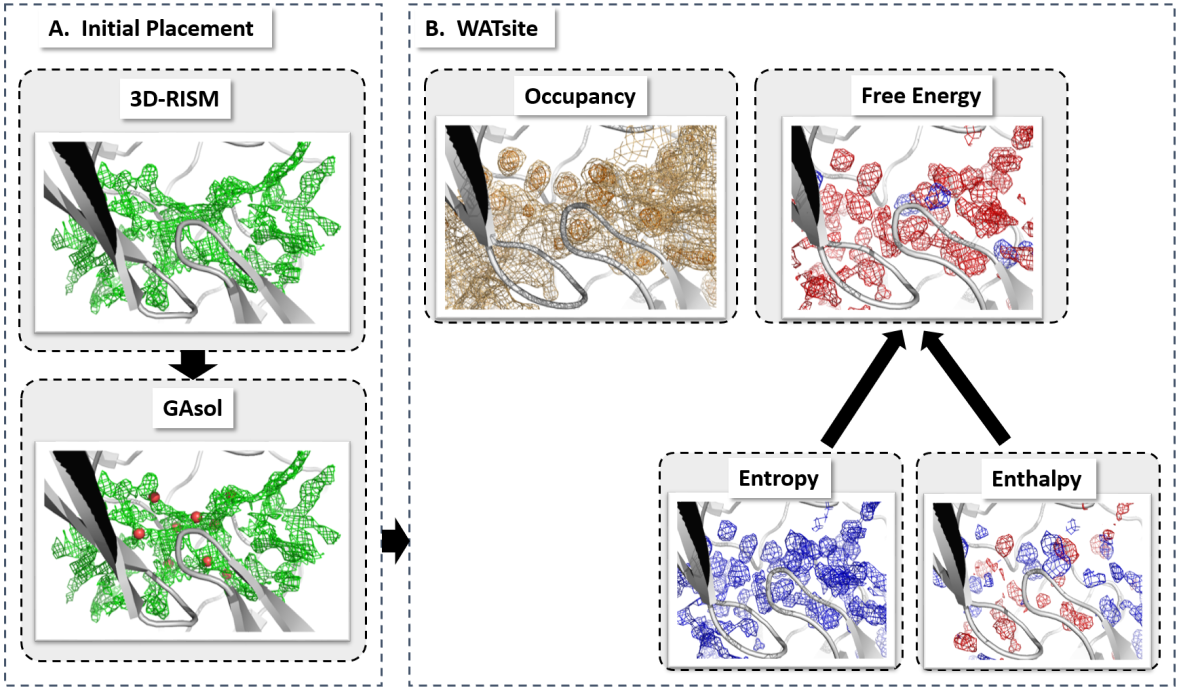}
  \caption{Overall procedure of WATsite combining (A) initial placement of water molecules using 3D-RISM and GAsol, and (B) subsequent MD simulation with explicit water molecules and WATsite analysis to generate water occupancy, enthalpy and entropy grids (adapted from \cite{Lill2019}).}
  \label{fig:Overview2}
\end{figure*}

Here, hydration site data was generated for several thousand protein systems using WATsite. The recently published protocol combining 3D-RISM, GAsol and WATsite (Figure \ref{fig:Overview2}) was used to achieve convergence for hydration site occupancy and thermodynamics predictions for solvent-exposed and occluded binding sites \cite{Masters2018}. Using 3D-RISM site-distribution function \cite{Kovalenko1998,Sindhikara2012,Sindhikara2013} and GAsol \cite{Fusani2018} for initial placement of water molecules, WATsite then performs explicit water MD simulations of each protein. Finally, explicit water occupancy and free energy profiles of each hydration site (=high water-occupancy spot) in the binding site are computed. This hydration data is distributed on a 3D grid encompassing the binding site and used as output layers for the neural networks to be trained on.

WATsite and other related hydration site methods rely on time-consuming simulations. A significantly more efficient method for hydration profiling is highly desirable, to utilize this concept on flexible proteins and large set of ligands with alternative binding poses, for example in drug discovery applications such as virtual screening to dynamic and flexible protein entities.

\subsection{Neural networks for WATsite prediction}
\begin{figure*}
  \centering
  \includegraphics[width=16.5cm]{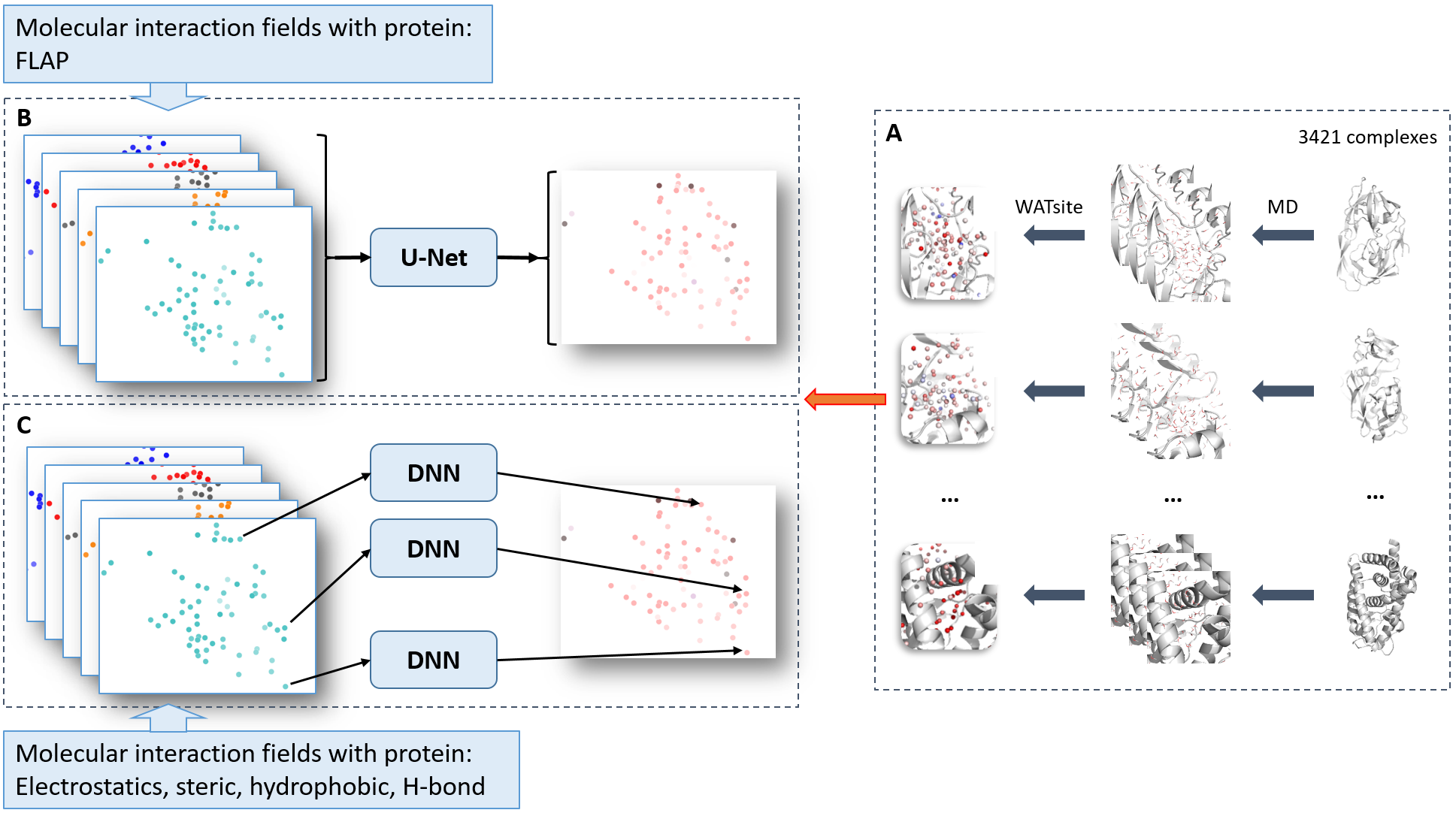}
  \caption{Overall procedure of prediction of WATsite data using neural networks. (A) Hydration data generation using WATsite simulation. Data is used as output layer for training of neural networks. (B) Direct prediction of complete 3D hydration image using U-Net approach. (C) Point-wise prediction using simple fully-connected neural network.}
  \label{fig:Overview1}
\end{figure*}

Two different types of neural networks have been designed to predict hydration information (Figure \ref{fig:Overview1} A). In both approaches, input descriptors were generated for each grid point representing the spatial and physicochemical environment of that potential water location. In the first approach, the complete 3D input grid was translated into a 3D output grid representing the hydration information using semantic segmentation approach (Figure \ref{fig:Overview1} B). In the second approach, the hydration information of each individual point is predicted based on the input descriptors (Figure \ref{fig:Overview1} C).

\subsubsection{Neural networks for semantic segmentation}
In the first approach to predict hydration data, we adapted deep neural network concepts commonly used in semantic image segmentation. Semantic image segmentation is the task to identify the pixels in an image that belong to a specific class or category, for example a specific object in an image. The great advantage of such networks is that they are able to be trained end-to-end by creating a mapping from the input layers to the output images. The resulting output is an image or a grid with the same dimensions as the input layers. 
Among the various architectures used for this task, U-Net has been demonstrated to often produce superior segmentation performance with smaller training sets compared to other methods \citep{unet}. 
As mentioned above, the typical task in segmentation is determining if a pixel is part of an object of interest or not. Here, we extended the segmentation task to multi-class segmentation estimating the occupancy of water molecules above multiple threshold values in different moieties along the protein surface.

\paragraph{Generation of descriptors.}
We used the "refined set v.2016" from the PDBind database \citep{pdbind1,pdbind2} consisting of 4057 protein-ligand complexes. Hydration site data was generated using WATsite as described in \citep{Lill2019}. 
The ligands were removed from the binding site for WATsite calculations but used to define the center of the hydration grids aligning them with the ligand centroids in the x-ray structure. 
The input descriptor grids were generated using FLAP  \citep{flap1,flap2}. The descriptor grids generated by FLAP were interpolated and aligned to the WATsite grids using MDAnalysis package \citep{mdanalysis1,mdanalysis2}. The process for selecting relevant chemical probes for FLAP is further explained in Section \textit{Probe selection}. FLAP occasionally failed to generate output for one or two probes for some proteins due to an internal program issue. As this is a commercial software, it was not possible to correct this error. PDB (Protein Data Bank) files for which FLAP failed to generate an output were removed. Finally, 3421 PDBs were used for training and testing of the neural network models.

\paragraph{Probe selection.}
In FLAP 78 probes could be used to generate interaction grids between protein and chemical probe. To reduce the number of input layers for the CNN model, we performed k-means clustering of the FLAP grids of three randomly selected protein systems. The distance matrix used during clustering was based on Pearson correlation coefficients between the interaction values on the 3D FLAP grids of a pair of probes. 
In detail, the distance between two interaction probe types was defined by subtracting the Pearson coefficient value from one. The number of clusters was chosen to be 12. One representative probe type from each cluster was used to finally generate as set of 12 representative probes with lowest correlations between their interaction grids. These grids represent 12 input channels to the neural network. Increasing the number of channels (probe types) did not show significant improvement for the network and only increased the training time.  

\paragraph{Processing of hydration occupancy data.}
Initially, the generated neural network models were designed to generate regression models to predict continuous occupancy values. These models, however, failed due to significant imbalance between low and high occupancy values (Supporting Information Figure S1). 
Alternatively, we proceeded with a multi-class segmentation model with six output channels. Each of those channels represents the water occupancy above a chosen threshold. In detail, WATsite occupancy values were transformed into labels based on the threshold values that were selected for the network. The threshold values were 0, 0.02, 0.03, 0.045, 0.06 and 0.07. 
Input data grids from FLAP were clipped at -20 and 20 kcal/mol and scaled to be within -1 and 1, to remove the rare very large or very small values. This range covers more than 99 \% of all points (Supporting Information Figure S2).

\paragraph{Network architecture and model building.}
Our neural network architecture was based on the work in \citep{segarch}, with the difference that in our implementation, the network contained six output channels. 
In detail, a modified version of a U-Net neural network was used which contains Residual connections and Inception blocks.
Residual connections were first introduced in ResNets \citep{resnet}. They have the advantage of preserving the gradient throughout a deep neural network addressing the vanishing gradient problem of those networks. 

Another issue is the optimization of the kernel size of the convolutional filters. Sub-optimal kernel sizes can lead to overfitting or underfitting of the network. Inception blocks have been designed to overcome this issue, where the Inception blocks contain convolutional layers with different kernel sizes running in parallel. Throughout the training process, the network learns to use the layers with convolutional kernel size that best fits the input data which results in better training process \citep{inception}.

\begin{figure*}
  \centering
  \includegraphics[width=16.5cm]{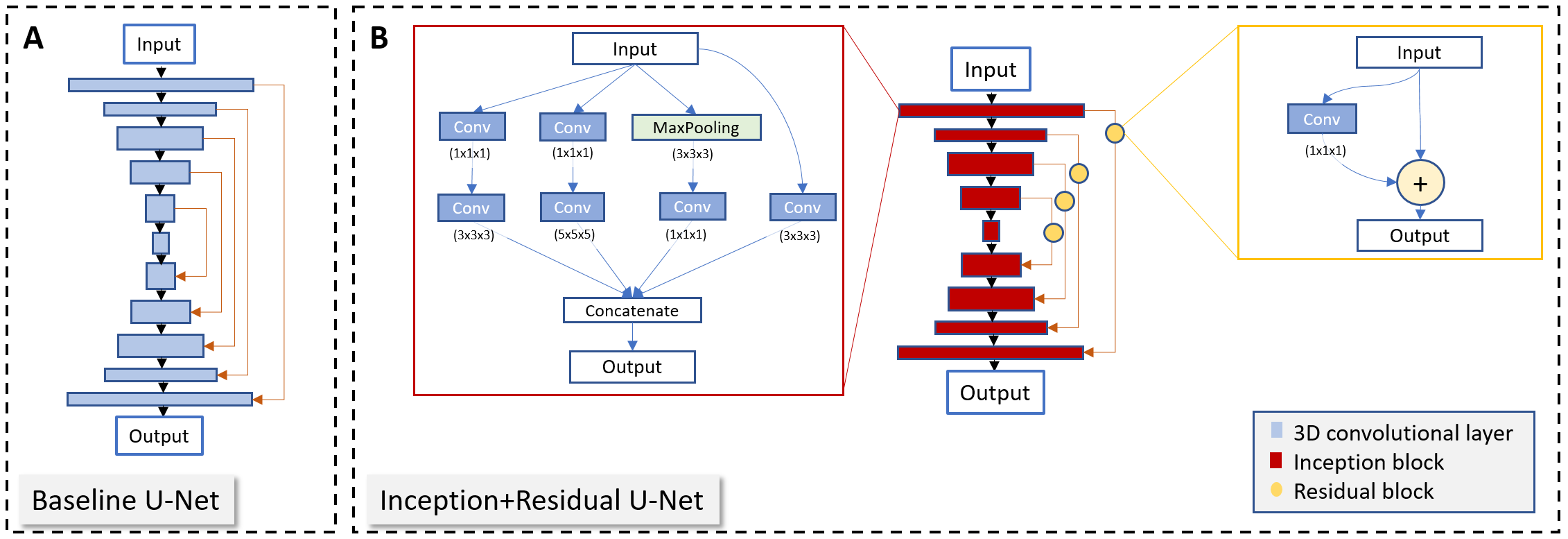}
  \caption{(A) Baseline U-Net and (B) Inception+Residual U-Net architecture used for multi-classification model for hydration density prediction.}
  \label{fig:Unet_architecture}
\end{figure*}

The U-Net that we used as a baseline model for our experiments consists of 6 encoder and 5 decoder layers (Figure \ref{fig:Unet_architecture} A). Each layer has a 3D convolutional layer with kernel size 2, stride size of 2 and zero padding. The number of filters for layers 1-6 is 32, 64, 128, 256, 512, and 512, respectively. Each convolutional layer was followed by a Batch Normalization layer, a Dropout layer and LeakyReLU activation. Each decoding layer consists of an Upsampling3D layer with size 2 followed by a convolutional layer, Batch Normalization layer, Dropout, and concatenation layer (which provided the skip connections in the U-Net) and ReLU activation. The number of filters for layers 7-10 is 512, 256, 128, and 64, respectively. The last layer consists of six filters (for the classification of 6 thresholds).

The Inception+Residual U-Net that we used resembles a U-Net, with the exception that each convolutional layer is replaced by an Inception block and the skip-connections contain a Residual block (Figure \ref{fig:Unet_architecture} B). Inception and Residual blocks and convolutional layers are followed by ReLU activation. The network has 5 encoder layers and 4 decoder layers. All Inception blocks are followed by a Dropout layer. Each decoder layer has an Upsampling3D layer prior to the Inception block. The last layer is a convolutional layer with filter number of 6 and kernel size 1.

As discussed above, regions in the grid with high water occupancy are sparse by nature, resembling a significant imbalance between number of low-occupancy and high-occupancy grid points. This makes the prediction of higher occupancy grid points difficult, as commonly used loss functions such as mean squared error will not work properly for such imbalanced data. The sparsity of the dense regions causes the network to predict low or zero values for all grid points even for high occupancy points. This problem also occurs in image segmentation tasks, where the object of interest is small compared to the whole image being analyzed, for example in the detection of small tumors in brain images \citep{gdl}. One of the loss functions that has been designed to train such imbalanced data is the Dice loss, which is a modified, differentiable form of the Dice coefficient \citep{vnet}. We used the generalized form of the Dice loss (GDL) \citep{gdl} which assigns higher weights to the sparser points: 

\begin{center}
$GDL=1-2\frac{\sum_{l=1}^{6}{w_{l}\sum_{n}\ r_{ln}p_{ln}}}{{\sum_{l=1}^{6}w_{l}\sum_{n}\ (r_{ln}+p_{ln})}}$\\
\end{center}
with label weights $w_l = 1/(\sum_{n=1}^{N}r_{ln})^2$ proportional to the inverse of their populations squared. $r_{ln}$ and $p_{ln}$ are the reference and predicted label (l) values at a grid point $n$, respectively \citep{gdl_weight}.
This loss function will strongly penalize sparse grid points, enforcing the learning algorithm to more precisely predict those values in addition to the large number of low-occupancy grid points. 

Adam optimizer \citep{adam} with learning rate of 0.001 and a batch size of 16 was used for training the model. Learning was performed for 100 epochs using Keras \citep{keras} with Tensorflow \citep{tensorflow2015-whitepaper} back-end. Once trained, the six output channels of the network are combined to obtain a grid with a range of values which represent the likeliness of hydration.

\subsubsection{Neural networks for point-wise prediction using spherical harmonics expansion}

In the second approach, the hydration information of each individual point is predicted based on the input descriptors specifying water-protein interactions at this location and the environment of this water location. The approach consists of two subsequent models, a classifier to separate grid point with water occupancy from those without, and a second regression model for occupied grid points to compute occupancy values and free energies of desolvation.

\paragraph{Classification model to identify grid points with water occupancy.}
For each grid point, the spatial environment and flexibility of surrounding atoms is computed. In detail, the distance from grid point $k$ to all atoms $i$ in the neighborhood of the grid point are computed and the van der Waals radius of the protein atom $\sigma_{i}$ is subtracted:
\begin{equation}
\label{equ:steric1}
\widetilde{r}_{ik} = |R_i - r_k| - \sigma_{i}. 
\end{equation}
All $\widetilde{r}_{ik}$ values up to 6 \AA\ are distributed onto a continuous 25-dimensional vector using the Gaussian distribution function, where the value at bin $i$ is 
\begin{equation}
\label{equ:Gaussian1}
p_{k,i} = \exp{\left(-\left(\widetilde{r}_{ik} - \left(i \cdot w - 1 \text{ \AA} \right)\right)^2/(2 \cdot w^2)\right)} 
\end{equation}
with $w=7 \text{ \AA}/25$. All values are finally scaled using $\tanh{(p_{k,i}/5)}$ to limit values to the range [0;1].

Separate vectors are computed in the same manner for hydrogen-bond donor and acceptor atoms. The motivation for this additional descriptors are that shorter distances between water and hydrogen-bonding groups are observed compared to hydrophobic contacts.

Despite the applied harmonic restraints, dynamic fluctuations of the protein atoms is observable throughout the WATsite MD simulations. These fluctuations can have impact on the accessibility of water molecules to different locations in the binding site. To incorporate those atomic fluctuations in the neural network predictions of occupancy, we designed a simple flexibility descriptor for the side-chain atoms (backbone atoms are considered rigid in this analysis). The shortest topological distance $t_i$ of a side-chain atom $i$ to the corresponding C{$_\alpha$} atom is translated using $f_i = 2 \cdot \tanh{(t_i/4)}$. The distance between this atom and grid point $k$ is then distributed to an additional 25-dimensional vector using a modified Gaussian distribution
\begin{equation}
\label{equ:Gaussian2}
q_{k,i} = f_i \cdot \exp{\left(-\left(\widetilde{r}_{ik} - \left(i \cdot w - 1 \text{ \AA} \right)\right)^2/(2 \cdot w^2)\right)} 
\end{equation}
Subtracting this vector $q_{k,i}$ from the unmodified vector $p_{k,i}$ generates a vector measuring the flexibility of the environmental atoms around grid point $k$. 

All four vector are concatenated generating a 100-dimensional input vector to the neural network for classification.

In addition to the input layer, the neural network architecture consists of a fully-connected hidden layer with 1024 nodes with leaky-ReLU activation and dropout layer with dropout probability 0f 0.5, followed by a second fully-connected hidden layer with 512 nodes with leaky-ReLU activation and a final output layer with sigmoid activation to classify each grid point as either occupied (1) or unoccupied (0). A threshold occupancy value of $10^{-5}$ in the input was used to separate occupied from unoccupied grid points.

Adam optimizer \citep{adam} with learning rate of 0.001 and a batch size of 250 was used for training the model. Learning was performed for 50 epochs using Tensorflow \citep{tensorflow2015-whitepaper}. 

\paragraph{Regression model.}
For each grid point, first the direct interactions between water probe and protein atoms is computed. In detail, electrostatic fields of the protein atoms $i$ at location $R_i$ with partial charge $Q_i$ are computed on each grid point $r_k$ 
\begin{equation}
\label{equ:electrostatics}
E^{elst}_k = \sum_{i} \frac{Q_i}{|R_i - r_k|}. 
\end{equation}
Steric contacts of water probe with protein atoms $i$ at location $R_i$ with van der Waals radius $\sigma_i$ and well-depth $\epsilon_i$ is computed using a soft alternative of the van der Waals equation
\begin{equation}
\label{equ:vdw}
E^{sterics}_k = \sum_{i} \sqrt{\epsilon_i \epsilon_p}\left(\left(\frac{\sigma_{ip}}{|R_i - r_k|}\right)^4-\left(\frac{\sigma_{ip}}{|R_i - r_k|}\right)^2\right). 
\end{equation}
with $\sigma_{ip}=\sigma_{i}+\sigma_{p}$ (probe $\sigma_{p}=1.6$ \AA) and  well-depth of probe $\epsilon_p=0.012$ kcal/mol. 
Protein parameters from the Amber14 force field are used.

Hydrophobic contacts are computed \cite{Li2011} using
\begin{equation}
    \label{eq:hphob}
       E^{hphob}_k = \sum_{i}%
   \begin{cases}
     1 &\text{if $s \le -1$} \\
     0.25 \cdot s^3 - 0.75 \cdot s + 0.5 &\text{if $-1 < s < 1$} \\
     0 &\text{if $1 \le s$}.
   \end{cases}
\end{equation}
with
\begin{equation}
\label{equ:hphob2}
s = 2.0 \cdot \left({|R_i - r_k|} - \sigma_{ip}-2.0\right)/3.0. 
\end{equation}

Hydrogen-bond interactions between water probe and protein acceptor/donor heavy atoms $i$ are computed using
\begin{equation}
\label{equ:HbondA}
E^{HBond-Acc}_k = \sum_{i} \exp{\left(-|R_i - r_k - R^0|^2\right)} 
\end{equation}
and
\begin{equation}
\label{equ:HbondD}
E^{HBond-Don}_k = \sum_{i}%
    \begin{cases}
        -\exp{\left(-|R_i - r_k - R^0|^2\right)\cdot\cos\left(\angle iHk \right) } & \text{if $\cos{\left(\angle iHk\right)} < 0$} \\
        0 & \text{if $\cos{\left(\angle iHk\right)} \ge 0$}
    \end{cases}
\end{equation}
respectively ($R^0 = 1.94$ \AA).

Each interaction term is then scaled and transformed using a hyperbolic tangent function to the range $[0;1]$ 
\begin{equation}
\label{equ:tanh}
\widetilde{E}^{property}_k = \tanh(E^{property}_k) 
\end{equation}
with the exception of the electrostatic interaction term which is scaled to be within $[-1;1]$ (small negative van der Waals interaction values are clipped off at zero).
Each scaled interaction term is finally transformed into a continuous vector of size 20 using Gaussian distribution functions, where the value at each bin $i$ is determined by
\begin{equation}
\label{equ:Gaussian}
p^{property}_{k,i} = \exp{\left(-\left(\widetilde{E}^{property}_k - (i \cdot w + \min\left(\widetilde{E}^{property} \right)\right)^2/(2 \cdot w^2)\right)} 
\end{equation}
(bin width of w=2/20 and w=1/20 for electrostatic interactions and all other interactions, respectively).
The five 20-dimensional vectors are concatenated to generate a 100-dimensional input vector to the neural network.

\begin{figure*}
  \centering
  \includegraphics[width=16.5cm]{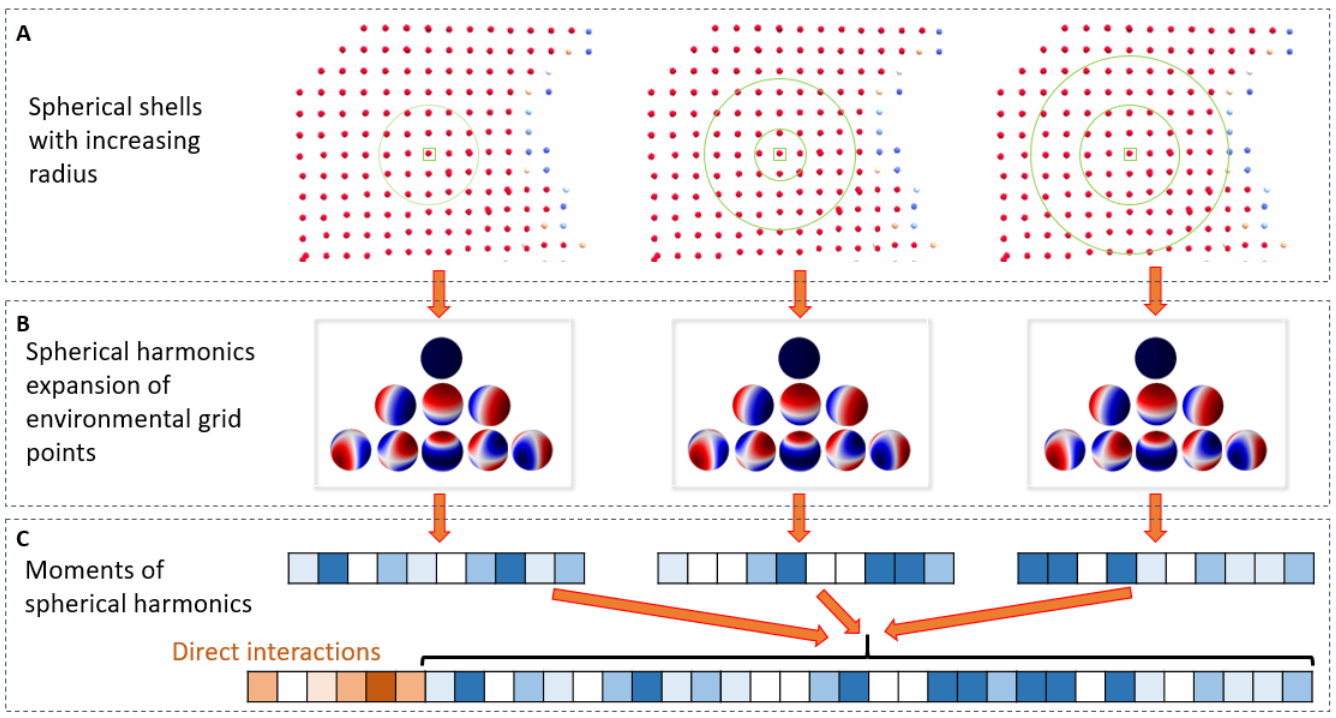}
  \caption{Generation of input vector for neural network for point-wise prediction of hydration data. (A) For each grid point, the interaction fields from the protein are computed. Nearby grid points within a spherical shell around the grid point are identified. (B) The interaction field distribution of those grid points are represented by spherical harmonics expansion. (C) The moments of this expansion generate an environment vector. (D) The environment vectors of spherical shells with increasing radius are concatenated together with the direct interaction fields at this grid point. This final vector is used as input for the neural network. }
  \label{fig:SphericalHarmonics}
\end{figure*}

The stability of water molecules not only depends on the protein environment but also on the surrounding network of additional water molecules. Thus, the environment of the water probe needs to be quantified as well. Here, we use a spherical harmonics expansion of the interaction fields on surrounding grid point as additional descriptors. 
In detail, seven spherical shells with increasing radius are defined to identify neighboring grid points with increasing distance to probe location: [$-\epsilon$; 1 \AA + $\epsilon$], [0.5 \AA -$\epsilon$; 1.5 \AA + $\epsilon$], $\ldots$, [3 \AA -$\epsilon$; 4 \AA + $\epsilon$] ($\epsilon$ is small value to include grid points with distance at the boundary of interval) (Figure \ref{fig:SphericalHarmonics}). The grid points in each shell are projected onto a unit sphere and the interaction values of those grid points are used to compute the coefficient of the spherical harmonics up to a certain order $l_{max}$:
\begin{equation}
\label{equ:spherical}
\widetilde{E}^{property}_{\text{neighbors of } k}(\theta,\phi) \approx \sum_{l=0}^{l_{max}} \sum_{m=-l}^l a_l^m Y_l^m (\theta,\phi)  
\end{equation}
The sum over the degrees of the L2-norm of the coefficients
\begin{equation}
\label{equ:spherical2}
\widetilde{a}_l = \sum_{m=-l}^l ||a_l^m||   
\end{equation}
is computed, transformed using $\tanh(\widetilde{a}_l)$ and distributed onto continuous 5-dimensional vectors using Gaussian distribution function (Equation \ref{equ:Gaussian}). The vectors of direct interactions (Equation \ref{equ:Gaussian}) are finally concatenated with the different coefficient vectors for the different $l$ and different interaction types to generate the final input vector to the neural network.

The neural network architecture consists in addition to the input layer a fully-connected hidden layer with 2048 nodes with leaky-ReLU activation and dropout layer with dropout probability 0f 0.5, followed by a second fully-connected hidden layer with 1024 nodes with leaky-ReLU activation and a final output layer with occupancy and free energy values. 

Adam optimizer \citep{adam} with learning rate of 0.001 and a batch size of 250 was used for training the model. Learning was performed for 125 epochs using Tensorflow \citep{tensorflow2015-whitepaper}.

\section{Results and Discussion}
\subsection{Neural network for semantic segmentation}
\paragraph{Failure of machine learning based on protein density descriptors.}
Initially, protein densities distributed on a 3D grid were used as input descriptors. Each atom is distributed on a 3D grid according to its atom type using a Gaussian distribution function centered on the atom center. Using this Gaussian smearing reduces the sparsity of the input data which would result in poor learning in neural networks since the gradients propagated throughout the network will be sparse as well \citep{waveforms}. Furthermore, Gaussian smearing better represents the spatial extension of the protein and therefore local accessibility of water to the protein surface.

Whereas these input data show good performance for binding pose prediction of chemicals binding to proteins \citep{gnina}, no significant learning was observed in the context of water occupancy prediction (data not shown). This failure can be interpreted by the lack of modeling of long-range protein-water interactions and water-water interactions. CNNs based on protein density will allow for modeling local correlation between protein shape/properties and adjacent water occupancy. The stability of water molecules in protein binding sites, however, is strongly influenced by long-range electrostatic interactions and by the formation of hydrogen-bonding water networks \citep{waternetworks,cavities}. Both contributions are difficult to model using localized features extracted by the layers of the CNN. 

To overcome this problem, we took a different approach in representing the protein structures to the CNN network, using molecular interaction fields (MIFs) data \citep{mifs}. MIFs are generated by first placing a fictitious probe molecule on each point of a 3D grid encompassing the binding site. The interaction value between probe and protein is rapidly calculated at each grid point assuming a rigid protein structure. Instead of providing an image of the protein, this approach rather generates a negative image of it and provides data for the binding site regions of the protein unoccupied by protein atoms but accessible to water molecules.
As described in Methods:Probe selection 12 probes were selected generating 12 different channels for the input layer.

\paragraph{Failure of point-to-point correlations using MIFs.}
Initially, neural networks were designed for simple point-to-point correlations, where the different MIF input channels were correlated with WATsite occupancy. In our tests, however, neural networks or even simpler machine learning algorithms were unsuccessful in finding any significant point-to-point correlations. From this observation, we concluded that MIFs, even the MIF generated with a water probe, differ significantly from the WATsite predictions. 
This can be explained by the fact that FLAP only analyzes direct protein-probe interactions and therefore lacks the incorporation of water-water interactions. Thus, the interaction value with a probe at a given point does not provide enough information for a network to infer water occupancy. 
For example, a grid point in an occluded space buried deep inside a protein may have a similar interaction profile with the protein in context of the MIFs than another grid point in a solvent exposed area. The former point, however, may have lower occupancy due to the lack of stabilizing water-water interactions.

WATsite in contrast includes water-water network interactions explicitly. Furthermore, it explicitly includes entropic contributions, as the water distribution is sampled from a canonical statistical ensemble during the MD simulation.
To predict water occupancy at a certain location, the neural network requires not only the interaction information on the corresponding grid point, but also the context of the grid point, i.e. interaction with other water-molecules. Those interaction can be represented either by directly including information of neighboring grid points or by the explicit design of input descriptors that include environmental information. 
The latter approach will be discussed in the following section "Neural networks for point-wise prediction using spherical harmonics expansion", the former will be discussed in the following paragraphs.
 
\paragraph{Performance in prediction of water occupancy grids.}
To incorporate the context of a grid point in the neural network, we utilized CNNs based on the computed MIFs. This approach predicts the water occupancy on a grid point by incorporating spatial context from surrounding grid points during the convolutional feature abstraction process. 
The CNN network architecture (Supporting Information S3) down-samples the input layer identifying features important for the prediction of water occupancy. The final layers up-sample the grid to the desired occupancy grid. Similar architectures have been used for many applications such as semantic segmentation and generative models.
More specifically, we use U-Net as the network architecture. U-Nets are commonly used for semantic segmentation tasks. For image segmentation tasks, a U-Net can rapidly learn to pass critical information such as the outlines of an object, which is similar between input and output layers. This process makes the learning more efficient. Similarly, for the task of water prediction, the surface of the protein is quickly captured by the U-Net from the input data. Our tests showed that without skip connections, it would be hard for the network to capture the protein surface, or the solvent accessible surface with the same efficiency. 

Initially, we attempted to generate regression models that aim to predict the actual occupancy value of each pixel or grid point. The resulting models showed poor prediction performance, which can be largely attributed to the highly imbalanced nature of the water grids, i.e. most grid points in a water grid have low or zero occupancy.  
Alternatively, the water prediction task using 3D CNNs can be tackled as a segmentation problem, detecting dense areas where water is more likely to have high occupancy. The problem of predicting water occupancy was formulated here as a multi-class segmentation problem allowing to identify regions with different levels of water occupancy, here predicting occupancy levels with threshold values of 0., 0.02, 0.03, 0.045, 0.06 and 0.07. 

For evaluating the neural network's performance, 5-fold cross-validation was used. The set of proteins was first divided into five groups. Then, the network was trained on four groups and tested on the one group left out, generating a set of five models. Given the similarity among the proteins in the refined set, we chose not to use random assignment to the five groups. For proper validation of the procedure, we instead minimized the similarity among the different groups by clustering the whole set of proteins based on binding site similarity. This guarantees that during cross-validation the test set is always least similar to the training set. To equalize the size of the clusters, samples were removed from larger clusters, resulting in 223 protein systems contained in each cluster. The similarity was calculated using the FuzCav program \citep{fuzcav} using the default configurations and the structures were clustered using the k-modes clustering algorithm \citep{kmodes1,kmodes2} on the feature vector generated by FuzCav. For the purpose of data augmentation, the training samples were rotated randomly on-the-fly along the coordinate axes. Prediction results for the cross-validation are shown in Table \ref{metrics-table}. Only data for the left-out systems are used in the statistical analysis.

\begin{table*}
  \caption{Various metrics for the performance of a baseline U-Net and a U-Net using Inception and Residual blocks. Performance on the validation sets are displayed (shown as mean $\pm$ standard deviation of cross-validation trials). Metrics are shown for the grids covering the whole binding site and for the sub-grids focusing on the area within 5 \AA\ of the ligand center. The results show that the Inception+Residual U-Net surpasses the baseline model's performance.}
  \label{metrics-table}
  \centering
  \begin{tabular}{|p{3cm}|p{2cm}|p{2cm}|p{2cm}|l|l|}
    \hline


    Network & Distance from Ligand     & Generalized Dice Loss       & Dice overlap (smoothed) \\
    \hline
    Baseline U-Net & Full grid  & 0.44 $\pm$ 0.08 & 0.40 $\pm$ 0.20  \\
    &$<$5 {\AA} from center & 0.35 $\pm$ 0.06 & 0.51 $\pm$ 0.17  \\
    \hline

    Inception+ & Full grid & 0.29 $\pm$ 0.04 & 0.79 $\pm$ 0.04  \\
    Residual U-Net         &$<$5 {\AA} from center & 0.24 $\pm$ 0.02 & 0.84 $\pm$ 0.02    \\
    \hline
  \end{tabular}
\end{table*}

\begin{table*}
  \caption{Precision and recall values for prediction of WATsite occupancy using fully convolutional neural network at five different levels of occupancy threshold values.}
  \label{regression_prec_rec_conv}
  \centering
  \begin{tabular}{|l|l|l|}
    \hline
    Occupancy threshold & Precision & Recall \\
    \hline
    0.02 & 0.86 $\pm$ 0.03 & 0.87 $\pm$ 0.03 \\
    \hline
    0.03 & 0.79 $\pm$ 0.02 & 0.81 $\pm$ 0.03 \\
    \hline
    0.045 & 0.73 $\pm$ 0.01 & 0.62 $\pm$ 0.03 \\
    \hline
    0.06 & 0.72 $\pm$ 0.01 & 0.56 $\pm$ 0.01 \\
    \hline
    0.07 & 0.70 $\pm$ 0.01 & 0.54 $\pm$ 0.00 \\
    \hline
  \end{tabular}
\end{table*}

\begin{figure*}
  \centering
  \includegraphics[width=0.9\linewidth]{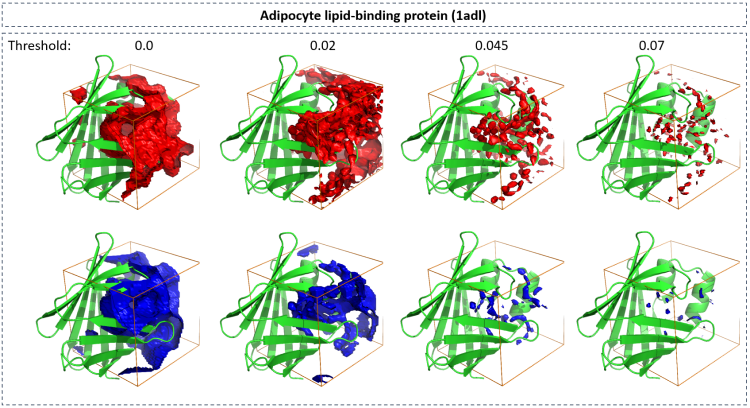}
  \includegraphics[width=0.9\linewidth]{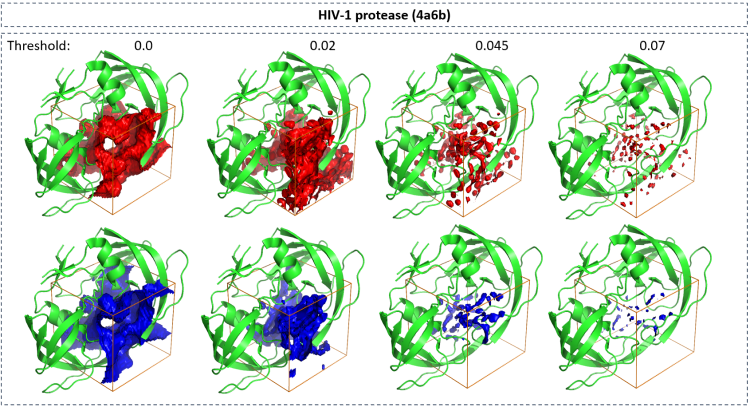}
  \caption{Visual comparison between ground truth (top row, red) and neural-network predicted (bottom row, blue) water occupancy for adipocyte lipid-binding protein (PDB-code: 1adl) and HIV-1 protease (4a6b). Predictions were performed using U-net. Isosurfaces at four different threshold values (0.0, 0.02, 0.045, and 0.07) are shown. The task of predicting areas with higher occupancy becomes challenging for the network due to the sparsity of those points (at thresholds 0.045 and 0.07). The regions closer to the corners of the grid are more difficult to predict as information of the context of those grid points is missing.}
  \label{fig:segm_whole}
\end{figure*}

Figure \ref{fig:segm_whole} shows visualization of the predicted water occupancy for two example proteins at different isovalues representing different thresholds of occupancy. At low thresholds, the quality of predicting occupancies is excellent; predicted and reference occupancy grids largely overlap. As the threshold is increased, the prediction quality drops due to the sparsity of the grid points with high occupancy, demonstrating that even with GDL the problem of imbalance in the data set was not completely resolved. 
We further observed, that the network fails to correctly predict the regions close to the boundaries of the grid. A possible explanation for this problem is that for these grid points the network does not receive the full context (MIFs of surrounding grid points) as those neighboring grid points would lie beyond the boundary of the grid box. This failure of predicting correctly the occupancy of boundary grid point, however, does not create a serious issue for the purpose of predicting hydration information in the binding site, as the grid points on the boundary of the box lie outside of the binding pocket volume. A mitigation for this problem is to remove the prediction in the boundary regions of the grid box after model generation. Therefore, we focused our analysis on the relevant region in the vicinity of the bound ligand, i.e. all grid points with a maximum distance of 5 \AA\ around the centroid of the co-crystalized ligand.

Table \ref{metrics-table} shows different metrics for the prediction quality of the model. The prediction performance for the test sets is shown. We used smoothed Dice overlap \citep{vnet} to measure the overlap between the reference and the predicted grids. In this metric the confidence of prediction of a label is included. For each metric both the quality for the full grid and for the area within 5 {\AA} from the ligand is displayed.

\begin{figure*}
  \centering
  \includegraphics[width=12cm]{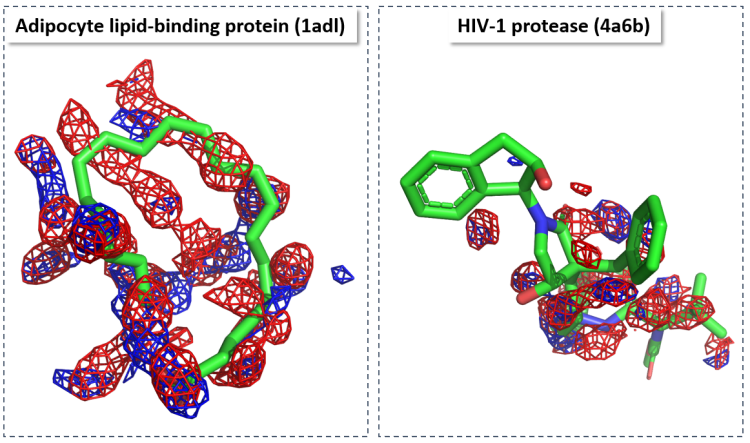}
  \caption{Visual comparison between group truth (top row, red) and neural-network predicted (bottom row, blue) water occupancy for adipocyte lipid-binding protein (PDB-code: 1adl) and HIV-1 protease (4a6b) within 5 \AA\ of the co-crystallized ligand. Note that the ligands were not included either in the water simulations to produce the ground truth or in the generation of input MIF grids. They were added for visualization purpose only. Predictions were performed using U-net. Isosurfaces at a threshold value of 0.045 are shown.}
  \label{fig:segm_lig}
\end{figure*}

Figure \ref{fig:segm_lig} shows an overlay of reference and predicted water occupancies within 5 {\AA} of the co-crystalized ligand to demonstrate the prediction quality in the proximity of the ligand. For applications of the model to drug design, we are interested in this particular region to identify how hydration might enhance, diminish or interfere with ligand binding at the binding site.

\paragraph{Importance of probes.}
We further analyzed which input MIF grids contributed most to the prediction performance. To compute feature importance we used Mean Decrease Accuracy (MDA) or permutation importance method \citep{Breiman2001}. This method measures how the absence of a feature decreases the performance of a trained estimator. This method can be directly applied on the validation set, without the need for retraining for each feature removal.
A feature is replaced with random noise with the same distribution as the original input. One simple way is to shuffle the values of a grid randomly, so that it no longer contains useful information. Not surprisingly, the probes which are most influential for the prediction quality were either water probes (OH2) or probes which mediate hydrogen bonding. It should be noted that although water probes from Flap are designed to indicate the water affine areas, they do not linearly correlate with WATsite occupancy, the Pearson correlation value between those MIFs and WATsite occupancy is close to zero. Table 2 shows the performance drop with shuffling of each input grid on the validation sets (sorted by importance of probe).
\begin{table*}
  \label{importance-table}
  \centering
    \caption{Dice overlap value for the cross-validation sets after shuffling of grid point value for each of the 12 MIF grids. The higher the decrease in value, the more important the probe grid is for the prediction. Important probe grids are displayed in bold. The un-shuffled dice overlap values are shown in Table 1 for all grid points and grid points around ligand.}
\begin{tabular}{|l|l|l|}
\hline
Probe & Dice overlap & Dice overlap (\textless{}5 {\AA} from ligand) \\
\hline
\textbf{C1=}   & 0.56 $\pm$ 0.06             & 0.58 $\pm$ 0.04                                        \\
\textbf{OH2}   & 0.51 $\pm$ 0.04              & 0.56 $\pm$ 0.03                                         \\
\textbf{CRY}   & 0.62 $\pm$ 0.04             & 0.67 $\pm$ 0.04                                          \\
\textbf{I}     & 0.63 $\pm$ 0.12              & 0.64 $\pm$ 0.09                                         \\
\textbf{O-}   & 0.71 $\pm$ 0.04             & 0.73 $\pm$ 0.02                                         \\
\hdashline
DRY   & 0.71 $\pm$ 0.07             & 0.79 $\pm$ 0.05                                          \\
N+    & 0.77 $\pm$ 0.04              & 0.83 $\pm$ 0.02                                          \\
H     & 0.75 $\pm$ 0.06             & 0.79 $\pm$ 0.02                                          \\
F3    & 0.78 $\pm$ 0.05             & 0.83 $\pm$ 0.03                                          \\
OC2   & 0.79 $\pm$ 0.05             & 0.84 $\pm$ 0.02                                          \\
I-H   & 0.78 $\pm$ 0.05             & 0.81 $\pm$ 0.02                                          \\
NA+   & 0.75 $\pm$ 0.03             & 0.81 $\pm$ 0.03
   \\
\hline 
\end{tabular}
\end{table*}

\subsection{Neural networks for point-wise prediction using spherical harmonics expansion}
\paragraph{Classification model.}
In contrast to the segmentation model, in the point-wise model each individual grid point represents a sample that can be used for training and testing of the model. Thus, the size of the data set is significantly increased and allowed to design a more aggressive testing protocol compared to the segmentation method.
For the point-wise prediction, the same 5-fold splitting procedure of the data set was used. On contrast to the segmentation model, only one-fifth was used for training and four-fifth for testing. 

\begin{figure*}
  \centering
  \includegraphics[width=10.5cm]{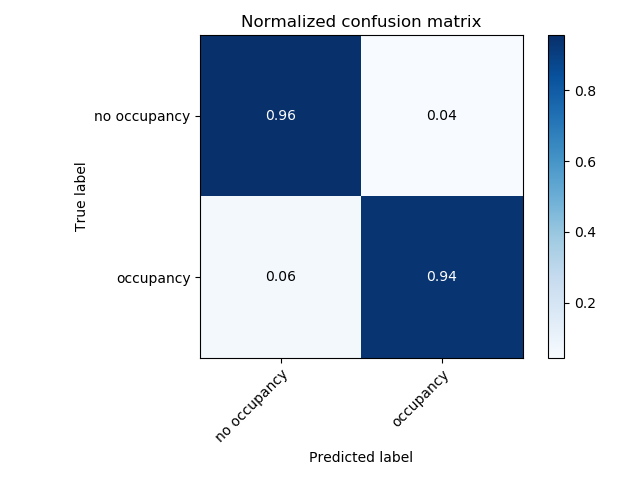}
  \caption{Normalized confusion matrix for classifying grid points with and without water occupancy using neural network model.}
  \label{fig:classif_test}
\end{figure*}

For the classification model, i.e. separating grid points with from those without water occupancy, the normalized confusion matrix over the test set was computed (Figure \ref{fig:classif_test}). 94\% of occupied grid points and 96\% of unoccupied grid points were correctly classified. The precision values of 0.97/0.92 and recall values of 0.96/0.94 for occupied/unoccupied data signifies the accuracy of the regression model in identifying moieties in the binding site that have observed water occupancy throughout WATsite simulations.

\paragraph{Regression model.}
Whereas the classification model allows to identify regions with likely water occupancy with high accuracy, a rather small occupancy threshold of $10^{-5}$ was used. In practice it is desirable to identify regions in the binding site with high water densities and occupancy peaks that resemble hydration sites. Therefore, a regression model was designed to identify those high density among low density regions.
Using descriptors encoding only the direct interactions with the protein at the specific grid point location (no inclusion of nearby grid points), a mediocre correlation between predicted and ground truth water occupancy was identified ($r = 0.52$) (Figure \ref{fig:regr_pred}).
Using only the radial distribution of interaction profiles of nearby grid points (l=0) increases the regression coefficient to  $r=0.82$. Increasing the depth of the spherical harmonics (l=1) only slightly increases the regression coefficient further to $r=0.85$. Further addition of angular functions to represent the environmental grid points (l=2) does not further improve the regression between ground truth and predicted occupancy values.
Consequently, we used the regression model with l=1 for subsequent analysis (see below):

\begin{figure*}
  \centering
  \includegraphics[width=7.5cm]{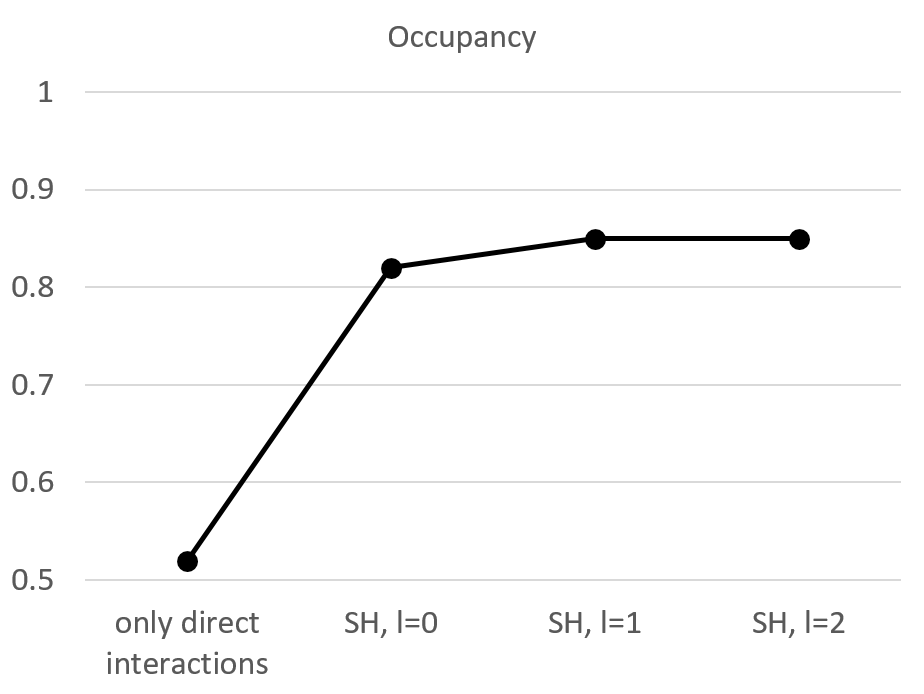}
  \includegraphics[width=7.5cm]{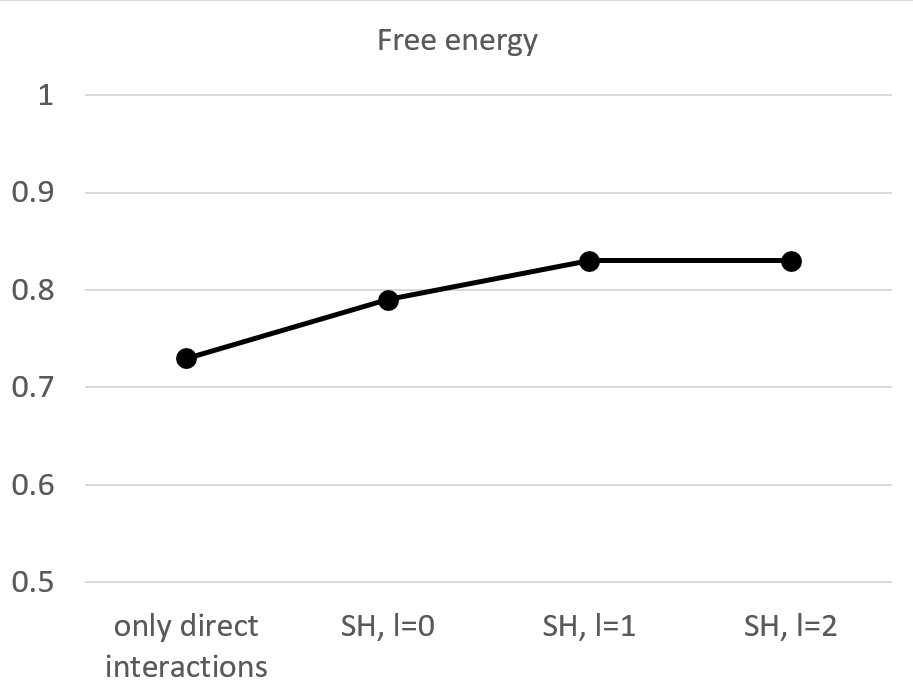}
  \caption{Regression coefficient $r$ for correlating occupancy and free energy values of neural network predictions with original WATsite data.}
  \label{fig:regr_pred}
\end{figure*}

The same trend, although weaker in magnitude was observed for the regression outcome for the free energy of desolvation at the grid points with occupancy. A maximum $r$ value of 0.83 was achieved.

For further evaluation of the neural network performance, 5-fold cross-validation was used. Again, only a fifth of the data set was used for training in each cross-validation step and four-fifth were used for testing the model. 
All five models generated very similar test set performance. For occupancy the $r$ values ranged between 0.85 and 0.86 (standard deviation of 0.004), for free energy it ranged between 0.83 and 0.84 (standard deviation of 0.0044). This highlights the robustness of the model, independent on the specific protein systems used for training.

\begin{figure*}
  \centering
  \includegraphics[width=15cm]{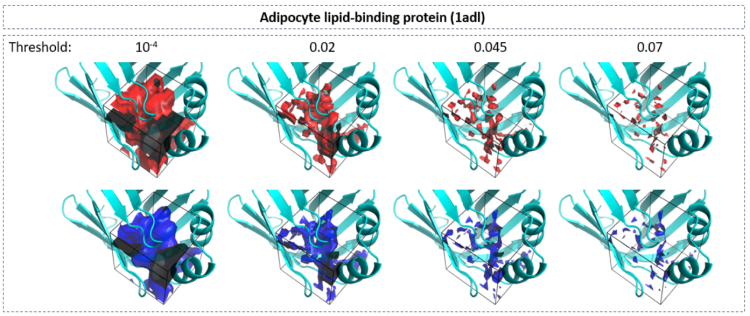}
  \includegraphics[width=15cm]{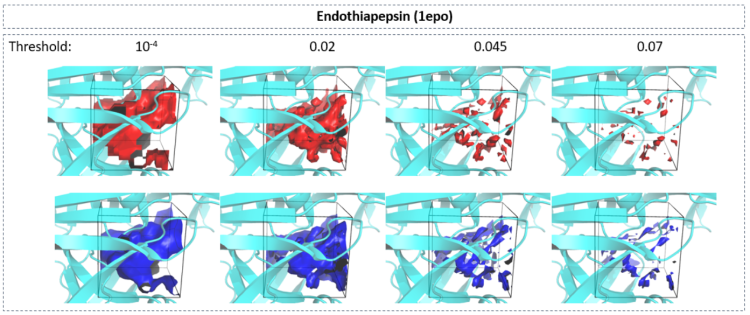}
  \caption{Visual comparison between group truth (top row, red) and neural-network predicted (bottom row, blue) water occupancy for adipocyte lipid-binding protein (PDB-code: 1adl) and endothiapepsin (1epo). Predictions were performed using regression neural network. Isosurfaces at four different occupancy values (10$^{-4}$, 0.02, 0.045, and 0.07) are shown.}
  \label{fig:example_occ}
\end{figure*}

Figure \ref{fig:example_occ} shows the comparison of predicted and ground truth water occupancy at isolevels of $10^{-4}$, 0.02, 0.045 and 0.07 for two different protein systems. 
Excellent overlap between predicted water occupancy and ground truth was observed with slight deterioration in accuracy for the highest density maps at 0.07. This visual observation can be quantified by measuring the precision and recall values at different classification threshold values of 0.02, 0.03, 0.045, 0.06 and 0.07 (Table \ref{regression_prec_rec}). Relative unchanged precision and recall values were observed up to an occupancy threshold of 0.045. Lower accuracy was observed for occupancy values of 0.06 and 0.07. This observation is consistent with previously discussed imbalance between large number of low-occupancy and small number of high-occupancy grid points.
\begin{table*}
  \caption{Precision and recall values for prediction of WATsite occupancy using regression neural network at five different levels of occupancy threshold values.}
  \label{regression_prec_rec}
  \centering
  \begin{tabular}{|l|l|l|}
    \hline
    Occupancy threshold & Precision & Recall \\
    \hline
    0.02 & 0.79 $\pm$ 0.03 & 0.79 $\pm$ 0.06 \\
    \hline
    0.03 & 0.79 $\pm$ 0.04 & 0.77 $\pm$ 0.06 \\
    \hline
    0.045 & 0.78 $\pm$ 0.04 & 0.76 $\pm$ 0.06 \\
    \hline
    0.06 & 0.75 $\pm$ 0.04 & 0.66 $\pm$ 0.06 \\
    \hline
    0.07 & 0.75 $\pm$ 0.04 & 0.66 $\pm$ 0.06 \\
    \hline
  \end{tabular}
\end{table*}

\begin{figure*}
  \centering
  \includegraphics[width=15cm]{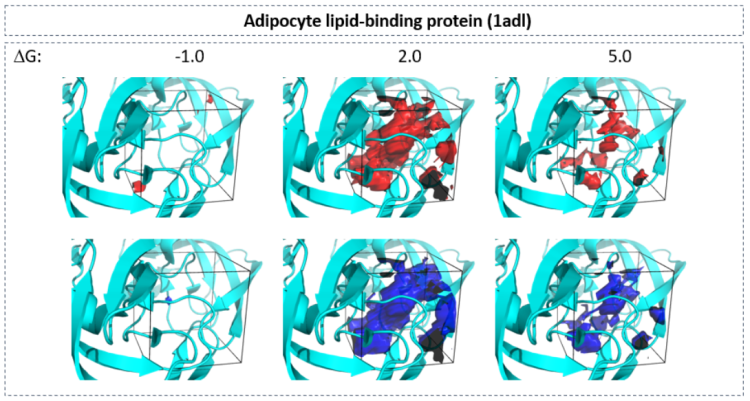}
  \includegraphics[width=15cm]{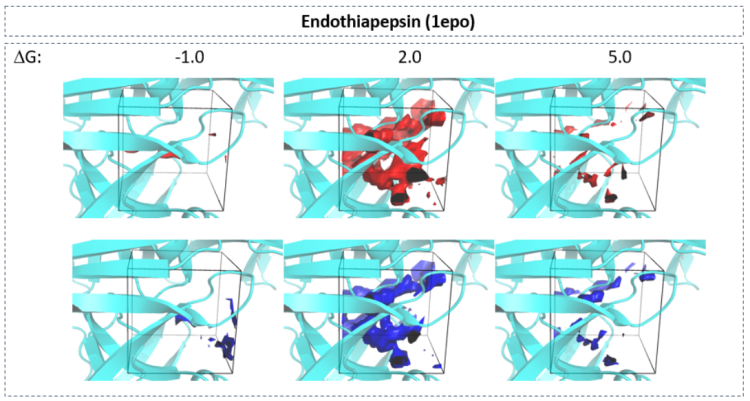}
  \caption{Visual comparison between group truth (top row, red) and neural-network predicted (bottom row, blue) desolvation free energy for adipocyte lipid-binding protein (PDB-code: 1adl) and endothiapepsin (1epo). Predictions were performed using regression neural network. Isosurfaces at three different free energy values (-1 kcal/mol, 2 kcal/mol, and 5 kcal/mol) are shown.}
  \label{fig:example_dG}
\end{figure*}

Similar trends were observed for the prediction of free energy values (Figure \ref{fig:example_dG}). Here infrequent negative desolvation values were less accurately predicted compared to positive values. Even regions containing high positive desolvation values were predicted with relative high quality.

\subsection{Applications}
\paragraph{Structure-activity relationships guided by hydration analysis.}
Hydration site prediction using MD-based methods such as WaterMAP or WATsite have been utilized in many recent medicinal chemistry projects for understanding ligand binding and structure-activity relationships(SAR), as well as for the guidance of lead optimization. Recently, Bucher et al. demonstrated the superiority of simulation-based water prediction usig WaterMAP compared to other commercial methods SZMAP, WaterFLAP and 3D-RISM \cite{Bucher2018} for the analysis on the structure-activity relationships of lead series of different target systems. To demonstrate the utility of neural-network predicted hydration information reproducing simulation-based hydration information in an efficient manner, we performed two retrospective SAR analyses on heat shock protein 90 (HSP90) and beta-secretase 1 (BACE-1).
\begin{figure*}
  \centering
  \includegraphics[width=15cm]{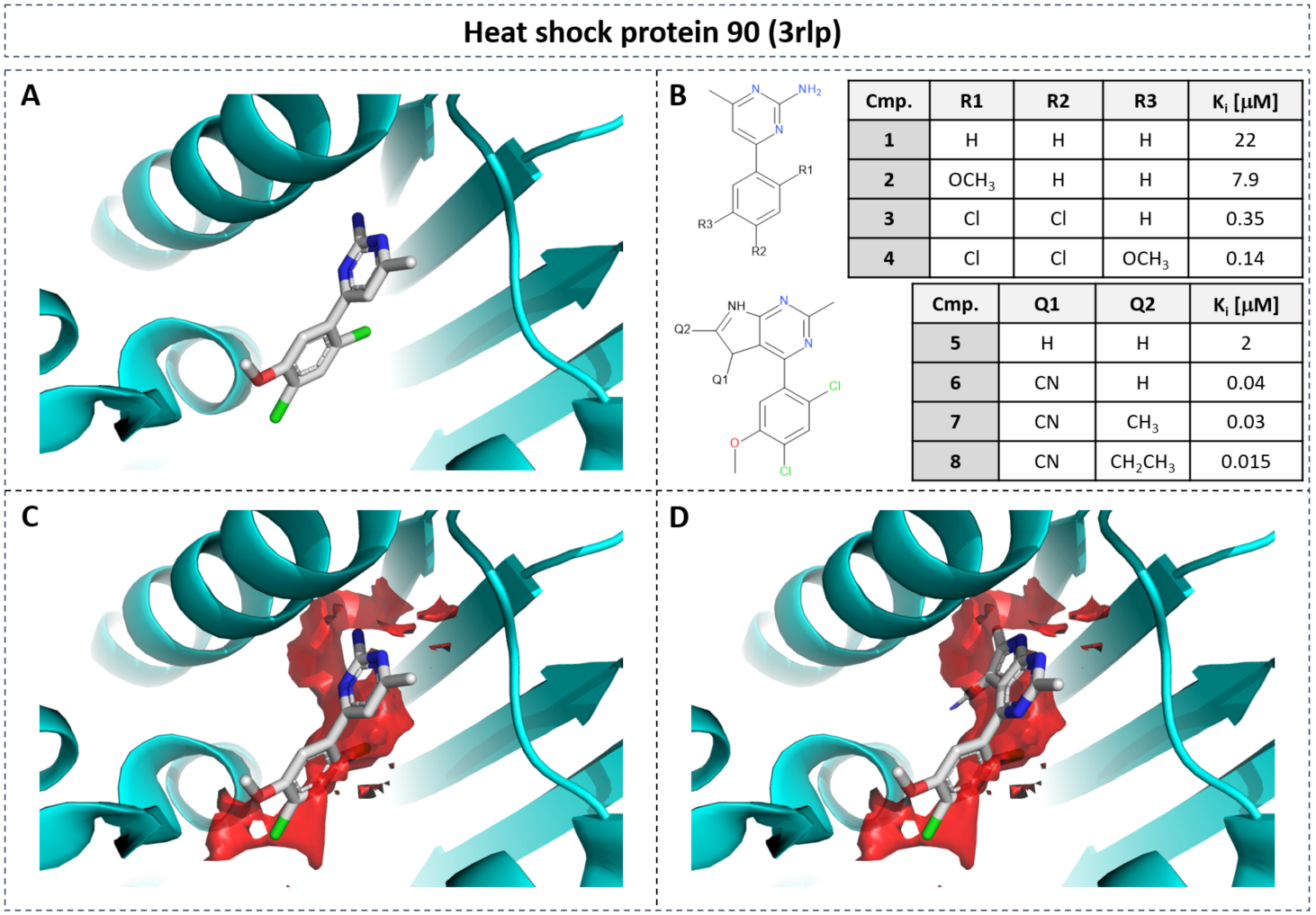}
  \caption{SAR of HSP90 inhibitors guided by gain in desolvation free energy based on point-wise neural network model. (A) Co-crystalized compound 4 in PDB structure with ID 3rlp. (B) SAR table of eight inhibitors with substituents replacing water density with unfavorable free energy (C/D: isolevel: 7.5 kcal/mol). (D) Compound 8 from X-ray structure 3rlr.}
  \label{fig:HSP90}
\end{figure*}
In a study of Kung et al.\cite{KUNG20113557}, a series of HSP90 inhibitors were synthesized and tested (Figure \ref{fig:HSP90}). The design of the molecules was guided by replacing water molecules resolved in the X-ray structure of HSP90. We performed hydration profiling on the x-ray structure 3rlp of HSP90 with the co-crystallized ligand removed using the point-wise neural network model. Water density with high positive (unfavorable) desolvation free energy (Figure \ref{fig:HSP90}C, red surface, isolevel for $\Delta$G=7.5 kcal/mol) is located around the phenyl ring of compound A  (Figure \ref{fig:HSP90}B). Subsequent substitution of hydrophobic groups on the phenyl ring at positions R1, R2 and R3 increases the affinity of the compound from 22 $\mu$M to 0.14 $\mu$M by replacing an increasing number of energetically unfavorable water molecules.
Additional water density with unfavorable free energy is located adjacent to the pyrimidine ring of the initial scaffold. Extending the pyrimidine scaffold to a pyrrolo-pyrimidine group and adding substituent at Q1 and Q2 position, replaced those additional unfavorable water molecules increasing the affinity by almost 10-fold to 15 nM.

\begin{figure*}
  \centering
  \includegraphics[width=15cm]{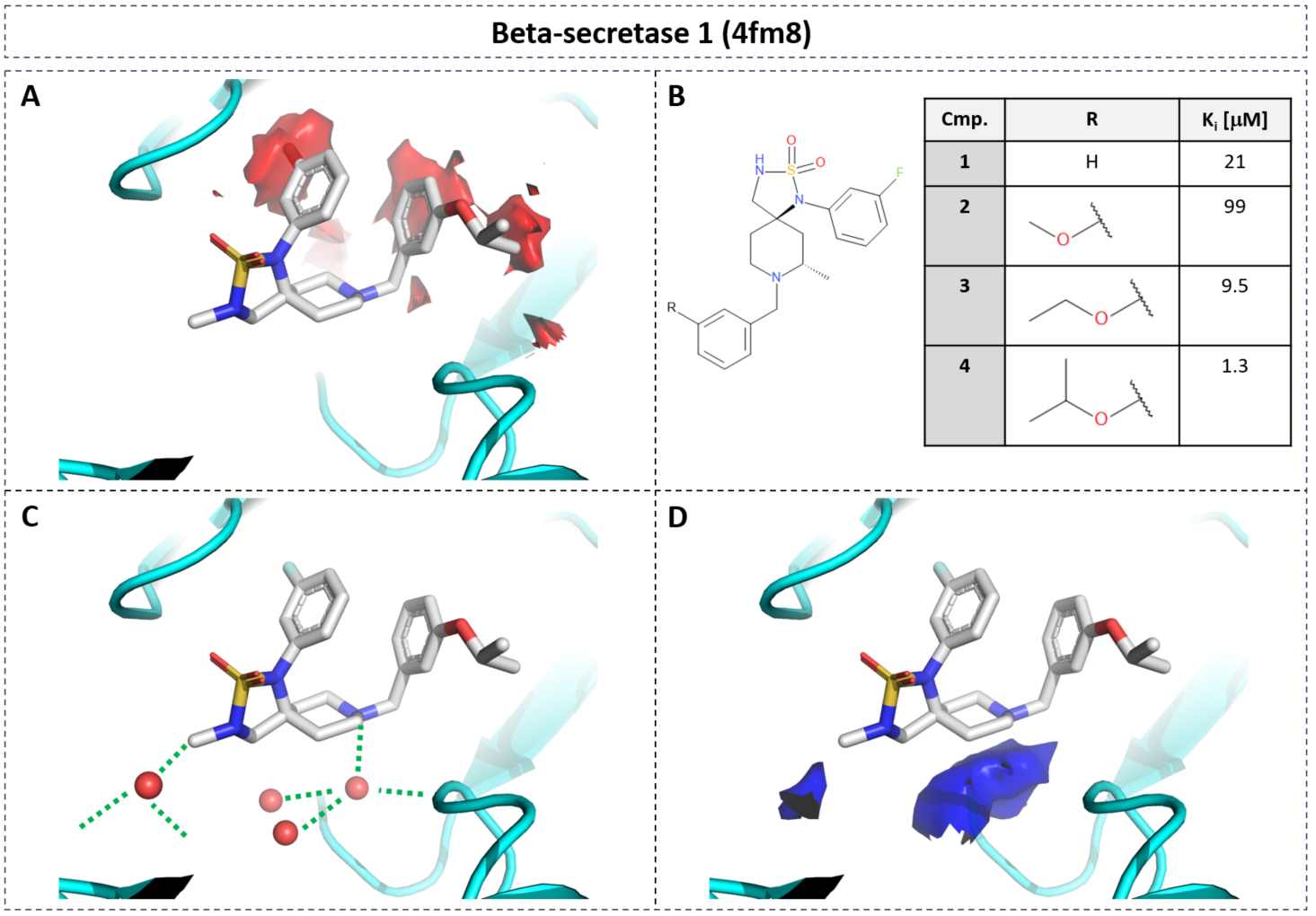}
  \caption{SAR of BACE-1 inhibitors guided by gain in desolvation free energy based on point-wise neural network model. (A) Co-crystalized compound 4 in PDB structure with ID 4fmp. (B) SAR table of four inhibitors with substituents replacing water density with unfavorable free energy (A: isolevel: 7.5 kcal/mol). (C) Water-mediated protein-ligand interactions overlap with water density with favorable enthalpy (D: isolevel: -3 kcal/mol).}
  \label{fig:BACE}
\end{figure*}
A similar retrospective analysis can be performed on beta-secretase 1 (Figure \ref{fig:BACE}). Focusing on the R-group of the terminal phenyl ring (Figure \ref{fig:BACE}B), density with unfavorable free energy is found adjacent to the R-group (Figure \ref{fig:BACE}A, red surface on the right). Methoxy substitution (Compound \textbf{2}) is not able to replace the water density, highlighted by a decrease in affinity. Elongated substituents such as O-ethyl (\textbf{3}) and O-isopropyl (\textbf{4}) spatially overlap with the unfavorable water density, replacing those water molecules. This results in significant affinity increase from 21 $\mu$M to 1.3 $\mu$M.
For BACE-1 two regions with favorable water enthalpy were observed (Figure \ref{fig:BACE}D, blue surface) that coincides with X-ray water molecules (Figure \ref{fig:BACE}C), mediating interactions between protein and ligand. Replacement of those water molecules should be considered with great care, as it may lead to decrease in binding affinity.

These two examples highlight the potential of our neural network approach to guide SAR-series expansion incorporating critical desolvation information ranging from the replacement of unfavorable water molecules to enthalpically favorable molecules mediating critical protein-ligand interactions.

\paragraph{Improved pose prediction}
In the second application we investigated if the reported performance of the neural network model (here the CNN model) can be utilized for improving ligand pose prediction. In a previous publication from our group \citep{Lill2019}, we showed significant improvement of pose prediction accuracy adding WATsite occupancy grids as additional input layers to a classification CNN model based on Gnina software \citep{gnina}. 
The major issue with this approach is that generating water occupancy grids for a large dataset of protein systems using WATsite or any MD-based water prediction program is computationally expensive. Here, the idea was to investigate if water grids generated via our CNN model can replace the data produced by WATsite to enhance the performance of Gnina.

In Gnina protein and ligand density are distributed on a 3D grid encompassing the binding site. For this distribution, a Gaussian distribution function centered on each heavy atom centroid is used. For each atomic element a separate distribution is computed for protein and ligand. This ensemble of occupancy grids is used as different channels of the input layer of a CNN that classifies native-like poses (RMSD $<2$ \AA) from decoy poses (RMSD $>4$ \AA).
Water occupancy grids predicted by our CNN model was used as additional input channel to the Gnina CNN.

To provide water occupancy data for Gnina, we retrained the water predictor network using 2288 and 1133 PDBs for training and test set, respectively. The train and test sets were based on the reduced set form Ragoza et al.\citep{gnina}. However, we increased the number of bad poses for a more realistic scenario. For each target protein, only one native-like pose with RMSD < 2 was chosen as good pose. We also made sure that every system contains a good pose. Systems with no good poses were removed reducing the data set to 1394 and 593 protein targets for training and test respectively. The training was performed for 10000 iterations. We used the default parameters and the reference model for pose prediction which is made available on Gnina's Github page (https://github.com/gnina/gnina). 

Here, we evaluated the performance of Gnina+water against Gnina alone and Vina/Smina. The results for Vina were obtained from Ragoza et al. \citep{gnina}. 

\begin{figure*}
  \centering
  \includegraphics[width=15cm]{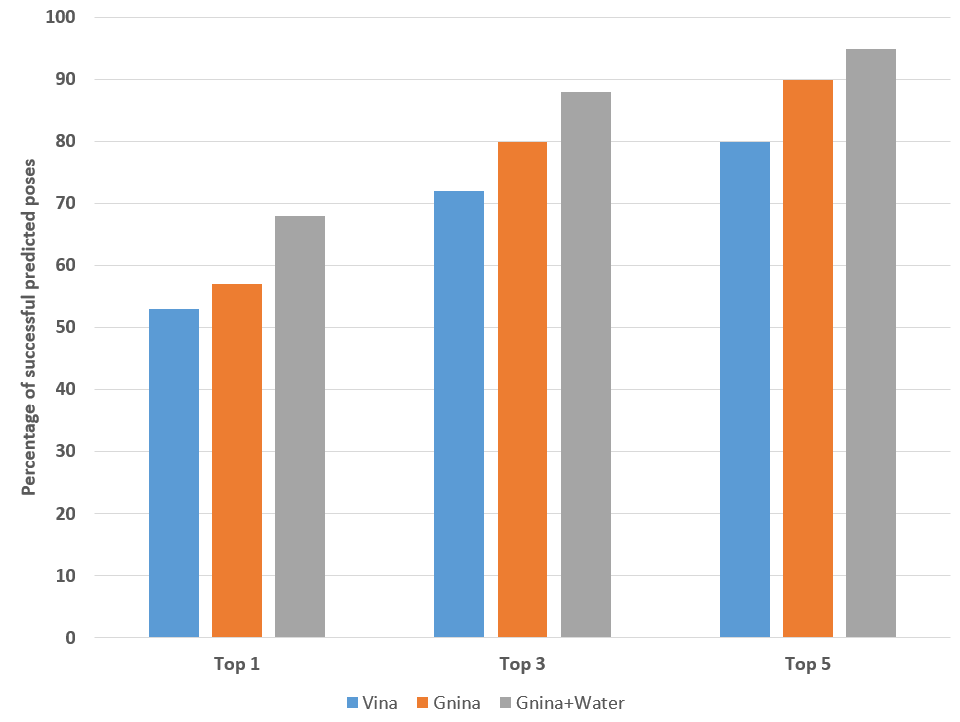}
  \caption{Percentage of protein systems with native pose (RMSD $<$ 2 \AA) in the test set within the top-1, top-3, and top-5 ranked poses using different scoring functions: Vina (blue), CNN with protein and ligand information (orange), and CNN with protein, ligand and WATsite occupancy information generated by U-net model (grey).}
  \label{fig:Gnina}
\end{figure*}

As it can be seen in Figure \ref{fig:Gnina}, inclusion of hydration occupancy from our neural network model into Gnina significantly increased the performance of Gnina on the test set. 




    

\section{Conclusion}
We presented the very first approaches to utilize neural networks for the on-the-fly prediction of thermodynamic hydration data obtained by time-consuming MD simulations.
Two alternative approaches were developed, one method that predicts the complete binding site hydration information in one network calculation in form of U-Net neural networks.  The other method relies on individually calculated descriptors for each grid point and point-wise predictions using neural networks based on fully-connected layers. 

In the former model, explicit regression was not feasible to obtain due to limited training data size and imbalanced occupancy values. Nevertheless, using multi-class classification and generalized Dice loss as a loss function, allowed the model to separate high from low density moieties in the binding site.

Using point-wise prediction in contrast, resulted in a significant increase in data size, reducing the effect of imbalance in data. This approach allowed us to generate good regression models for the prediction of occupancy and free energy of desolvation.

Application of the predicted hydration information to SAR analysis and docking, demonstrated the potential of this method for structure-based ligand design. Future applications include the marriage of protein flexibility and  desolvation data in ensemble docking, where precise hydration data could be computed for alternative protein structures, ligands and binding poses in sufficient computation time, a task until now unthinkable. This will allow the inclusion of explicit desolvation, water-mediated interactions and enthalpically stable hydration networks around the protein-ligand complex \cite{Lill2019}, a solvation component currently neglected in structure-based ligand design.



\bibliography{achemso-demo}

\newpage
\renewcommand{\thefigure}{S\arabic{figure}} 
\setcounter{figure}{0}.
\textbf{Supplementary Information}

\begin{figure*}
  \centering
  \includegraphics[width=15cm]{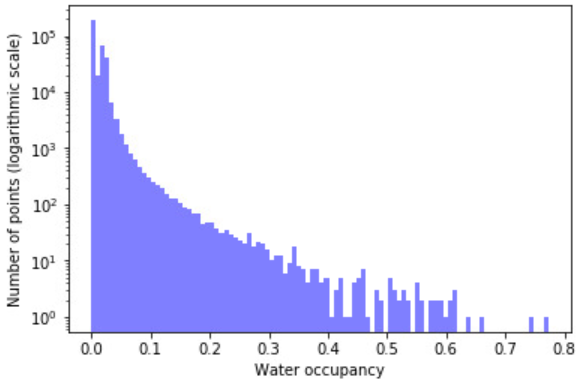}
  \caption{Number of grid points with corresponding WATsite occupancy values in selected systems.}
  \label{fig:histogram_watsite}
\end{figure*}

\begin{figure*}
  \centering
  \includegraphics[width=15cm]{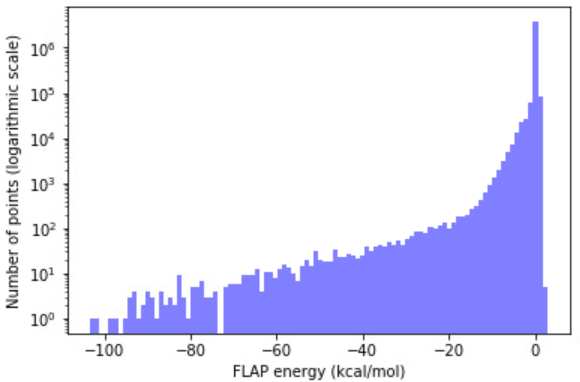}
  \caption{Number of grid points with corresponding FLAP energy values in selected systems.}
  \label{fig:histogram_watsite}
\end{figure*}

\begin{figure*}
  \centering
  \includegraphics[height=20cm]{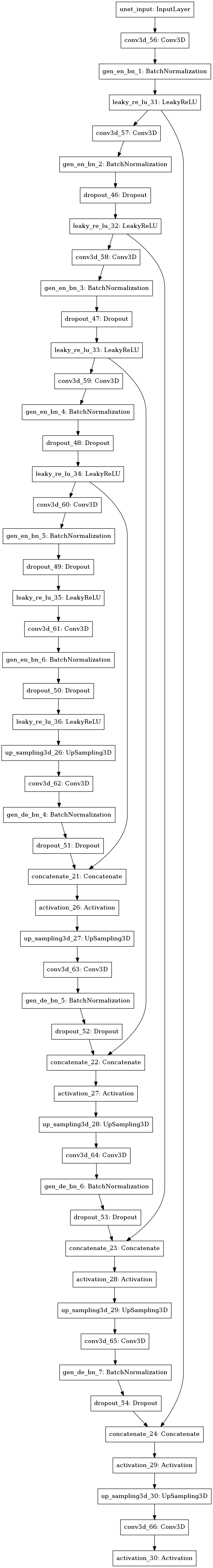}
  \includegraphics[height=20cm]{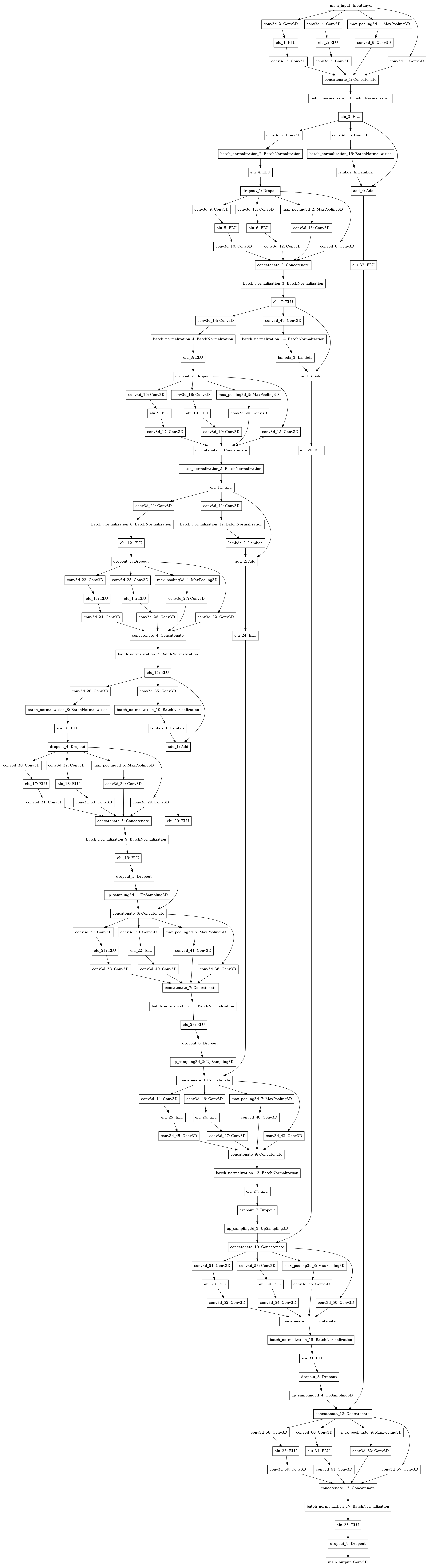}
  \caption{(A) Basic U-net architecture, and (B) Inception plus residual U-net architecture.}
  \label{fig:histogram_watsite}
\end{figure*}

\end{document}